\documentstyle[preprint,aps]{revtex}

\begin{document} 
\draft
\title{ Clustering, Order, and Collapse in a Driven Granular Monolayer}
\author{J. S. Olafsen \cite{jso} and J. S. Urbach}
\address{Department of Physics,Georgetown University,Washington, D.C. 20057}
\date{\today}
\maketitle

\begin{abstract}
Steady state dynamics of clustering, long range order, and inelastic collapse
are experimentally observed in vertically shaken granular monolayers.
At large vibration amplitudes, particle correlations show only
short range order like equilibrium 2D hard sphere gases.  
Lowering the amplitude ``cools'' the system, resulting in
a dramatic increase in correlations leading to either clustering or
an ordered state.  Further cooling forms a collapse: 
a condensate of motionless balls co-existing with a less dense gas.
Measured velocity distributions are non-Gaussian, showing nearly exponential 
tails.
\end{abstract}


Granular systems exhibit static and dynamic disorder \cite{JaegerNB96},
inviting a statistical description.  Results of
equilibrium statistical mechanics are not directly applicable to excited
granular systems, however, due to the large dissipation during collisions.  
In particular, the phenomena of clustering and collapse in a freely cooling,
initially homogeneous granular medium are examples of nonuniform energy 
distributions arising from instabilities in the collective motion as
energy is lost through interparticle collisions
\cite{GoldhirschZ93,McNamaraY96,NoijeEBO97,EsipovP97}. 
The instabilities lead to clusters, regions of high particle
density and increased dissipation rate.
The continued dissipation of energy in the freely cooling 
case eventually leads to inelastic collapse, where two or more particles 
lose all relative motion \cite{McNamaraY96}. 
By contrast, in an isochoric, equilibrium elastic hard disk system particle 
correlations do not increase as the temperature is decreased
\cite{LandauL}.

Recent work has extended the theoretical analysis to driven systems
\cite{TaguchiT95,Williams96,EuF97,GrossmanZB97,MullerLH98,NoijeE98}. 
The driven granular medium reaches a steady state where the energy lost 
through collisions is equal to the amount added externally.  By decreasing
the rate of energy input, the system may be slowly cooled, 
maintaining the dynamics near the steady state.
In this letter, we describe measurements of a large number 
(8,000--14,500) of spherical
particles on a vertically shaken, smooth horizontal plate.
Unlike recent investigations of thin vertical granular layers
\cite{ClementR91,WarrJH94} and a thin horizontal
layer driven from one side \cite{KudrolliWG97}, there are no large scale 
density gradients in this system, allowing for precise measurements
of the steady state statistical properties.

The particles used in this experiment are uniform 1 mm diameter stainless steel
balls \cite{expref} that constitute less than one layer coverage on a 20 cm 
diameter aluminum plate.  
As long as the shaking amplitude is not too large, the particles cannot hop 
over one another and the motion is effectively two-dimensional (2D); 
the energy imparted to the 
vertical component of the velocity by the plate is converted into horizontal 
motion through the particle-particle collisions.  The coefficients of 
restitution in this investigation are about 0.9 \cite{restit}.
 
In these experiments, the particle-particle collision rate 
is on the order of the shaking frequency, $\omega$.  The inter-particle 
collisions
destroy the delicate coherence necessary for the intermittancy observed
in the single sphere problem \cite{LuckM93}.  
Unlike thick granular layers where the collision rate is 
much higher than the shaking frequency \cite{MeloUS95}, particle-plate 
collisions are observed throughout the plate oscillation cycle.
For a vertical displacement in the shaking plate, 
$z(t) = A sin(\omega t)$, the dimensionless acceleration may be 
defined as $\Gamma \equiv A \omega^2/g$, where g is the acceleration due to 
gravity.  The acceleration is
uniform across the plate to within 0.2\%.  As the system is nearly 2D, 
cameras placed above the plate can capture the horizontal dynamics
of any particle in the system.  A high resolution camera is used to measure
spatial correlations in the monolayer \cite{highf} and a high speed camera
to obtain particle velocities between collisions \cite{stats}.

The motion is begun by shaking the plate at an acceleration where all of the 
particles are in motion in the gas-like phase, typically $\Gamma = 1.25$.  
The gas is characterized by an apparently random distribution of particle
positions and velocities.  Figure \ref{fig:one}(a) is an instantaneous image
of the gas for a population of 8000 particles (filling fraction 
$\rho = 0.463$ \cite{expref}) at $\Gamma = 1.01$ at a frequency of 
$\nu = \omega /(2 \pi) = 70$ Hz.  
The bright spots are reflections
on top of the particles; the particle diameter is a little more than twice the
diameter of a bright spot in the instantaneous images.
Figure \ref{fig:one}(b) is a time-averaged 
picture of 15 frames taken over a period of 1 s, and shows the
lack of any stable structure \cite{movie}.

As the acceleration amplitude is slowly decreased (at constant 
frequency), the average kinetic energy of the particles decreases, and
localized transient clusters of low velocity particles appear. 
The clusters break up over a timescale
of 1--20 s through interactions with higher velocity particles.
Figure \ref{fig:one}(c) is an instantaneous image 
when the acceleration is decreased to $\Gamma = 0.80$.  
To the eye, there is no significant difference from Fig. \ref{fig:one}(a), 
however in the time-averaged image, 
Fig. \ref{fig:one}(d), bright peaks are clearly evident, corresponding 
to low-velocity particles that have remained relatively close
to each other over the 1 s interval.  These 
clusters typically extend two to three ball diameters.

As the acceleration amplitude is further decreased to 
$\Gamma = 0.76$, the typical cluster size increases to 12--15 particles.  
Within a few minutes at this acceleration, one of these large clusters
will become a nucleation point for a collapse \cite{center}:  a condensate 
of particles that come to rest on the plate while in contact with one another.
The collapse reaches a steady state after several minutes, typically 
forming an ``island'' completely surrounded by a co-existing gas.
Figure \ref{fig:one}(e) is an instantaneous image of part of a typical 
collapse.  
The particles in the close-packed lattice remain in {\em constant} contact 
with each other and the plate.
Similar crystalline structure has been observed in a multi-layered system 
under horizontal shaking \cite{PouliquenNW97}; similar phase
separation has been seen in numerical simulations
\cite{TaguchiT95,Williams96}.
The time-averaged image in Fig. \ref{fig:one}(f) demonstrates that the 
particles in the collapse are stationary, while the particles in the 
co-existing gas phase are in constant motion.  
The two-phase co-existence persists essentially unchanged for as
long as the driving in maintained.

At higher densities (increased filling fractions), instead of a transition 
directly from clustering 
behavior to collapse, there is an intermediate phase with apparent long range 
order.  Unlike the collapse, the particles {\em do not} come to rest, but 
fluctuate about sites arranged on a hexagonal lattice.  In addition, there is
no phase boundary; the ordered phase extends across the entire cell.
Figs. \ref{fig:one}(g)-(h) show the behavior for N = 14,500 ($\rho = 0.839$)
and a drive frequency of 90 Hz.  The hexagonal lattice is apparent in 
Fig. \ref{fig:one}(g), although there is considerable positional disorder
and one vacancy.  
That disorder is almost completely absent in the time-averaged image,
Fig. \ref{fig:one}(h), demonstrating that the particles fluctuate
about sites on a regular lattice. 
The motion is visually similar to that seen in colloids \cite{Strandburg}.

Figure \ref{fig:two} is a ``phase'' diagram for two different densities.
The closed circles show the acceleration, $\Gamma$,
for which collapse forms as the gas is cooled as a function of frequency.
The system is hysteretic:  once the collapse nucleates, 
the driving amplitude must be increased a small amount, indicated by 
open circles in part (a) for N = 8000, to return all of the particles to the 
gas phase.
In part (b), results for N = 14,500 are shown (for clarity, 
the evaporation line is not shown).  
For frequencies below 70 Hz, the cooling gas undergoes a transition to a
collapse directly from a disordered phase.  At frequencies above
70 Hz, the medium first undergoes a transition to an ordered state 
(Figs. \ref{fig:one}(g)-(h)), indicated by the diamonds, and upon further 
cooling undergoes another transition to a collapse and co-existing gas-like 
phase.  Both the ordered and collapsed phases can be identified by eye.

The nature of the correlations in the gas phase are more subtle than in the
collapsed or ordered phases but can be quantified.  Particle
positions, determined from high resolution pictures \cite{highf}, are used to 
calculate the particle-particle correlation function, 
$G(r) =  <\rho(0) \rho(r)>/<\rho>^2$
where $\rho$ is the particle density.  In a hard sphere equilibrium gas,
$G(r)$ shows no significant correlations beyond one particle diameter.  
The correlations are due only to geometric factors of excluded volume
and are independent of temperature \cite{LandauL}.

The solid line in Fig. \ref{fig:three} shows $G(r)$ from a 
Monte Carlo calculation of a 2D gas of elastic hard disks in 
equilibrium for a density of 0.463 \cite{ChaeRR69}.
The experimentally measured correlation function in the gas-like phase 
($\Gamma = 0.892, \nu = 70$ Hz), shown by the open circles, is almost 
identical to the equilibrium result.  There are no free parameters in 
Fig. \ref{fig:three}.  The remarkable agreement clearly
demonstrates that the structure in the correlation function of the gas-like 
phase is dominated by excluded volume effects.
As the granular medium is cooled, the correlations grow significantly.
This is evident from the data for an acceleration of $\Gamma = 0.774$
($0.5\%$ above the acceleration where collapse forms), shown as filled 
diamonds in Fig. \ref{fig:three}.
The increased correlations indicate that there are non-uniform density 
distributions in the medium:  regions of high density that, due to the 
closed nature of the system, imply regions of low density.  

A crucial ingredient of a statistical approach to describing
the dynamics in a granular system is the velocity distribution,
which may show non-equilibrium effects as does the correlation
function.  
Velocity distributions that obey Maxwell statistics have been 
used in the formulation of many kinetic theories of granular systems
and the deviations due to inelasticity have been assumed small 
\cite{GoldhirschZ93,NoijeEBO97,EsipovP97,EuF97,NoijeE98,Campbell90,KnightW96}.
Recent results from simulations \cite{TaguchiT95,GrossmanZB97} and
experiments \cite{ClementR91,WarrJH94} demonstrate deviations from
Gaussian velocity distributions, but the experiments were not able to
resolve the functional form. 
With the use of a high speed camera \cite{stats}, 
the particle velocities can be determined between collisions.  
Extensive measurement of the velocity distributions in the plane of the
granular gas in our system demonstrate non-Gaussian behavior.

Figure \ref{fig:four} shows that experimentally measured
velocity probability distributions in the gas (circles), clustering (squares),
ordered (diamonds) phases, and the
gas phase coexisting with a collapse (triangles) all deviate significantly 
from a Maxwellian distribution (solid line).
The distributions are scaled only by a 
characteristic velocity $v_0 = (2 v_{2m}^2)^{1/2}$ proportional to the second 
moment of the distribution, $v_{2m}$, and approximately follow a universal 
curve.  
As no asymmetry is observed, the distributions contain both $v_x$ and $v_y$
data.  
The second moment of the distribution varied from  
3.8 cm/s in the gas phase ($\rho = 0.463$, $\Gamma = 1.01$, $\nu = 75$ Hz) to 
1.5 cm/s in the ordered phase at high density 
($\rho = 0.839$, $\Gamma =1.0$, $\nu = 90$ Hz).
The straight tails in Fig. \ref{fig:four}(b) suggest an exponential
velocity distribution.
This result is robust:  data acquired for sine, sawtooth, and square 
acceleration waveforms all demonstrate velocity distributions that 
scale onto one curve, indicating that the horizontal dynamics are dominated 
by particle-particle rather than particle-plate interactions.

The deviations from the Maxwell velocity distribution may result from
clustering in momentum space:  when the clusters form in real space they 
generate a local population of high-density, low-velocity particles
\cite{TaguchiT95}.  
The corresponding low density populations are regions of 
decreased dissipation that allow for high-velocity particles.  
Thus, as seen in Fig. \ref{fig:four}, the populations of low- and 
high-speed particles are higher than that expected from a Maxwell
distribution, and the intermediate velocity population is less.  
It is interesting to note that at the acceleration where the
correlation function looks like that of the equilibrium gas 
(Fig. \ref{fig:three}, open circles) the velocity distribution is strongly 
non-Gaussian.
Future work will investigate the behavior of the velocity distribution
at higher accelerations and 
the cross-correlation between the position and velocity
distributions.  A systematic picture of the microscopic statistics and the
macroscopic phases of this simple but rich system should provide a useful
testbed for theories of excited granular media.

The authors wish to acknowledge helpful conversations with Arshad Kudrolli 
and Jerry Gollub, and thank Harvey Gould for directing
us to Ref. \cite{ChaeRR69}.  This work was supported by grants from the 
Petroleum Research Foundation and the Sloan Foundation.

%
\begin{figure}
\caption{Instantaneous (left column) and time-averaged (right column) 
photographs detailing the different phases of the granular 
monolayer.  (a), (b), uniform particle distributions typical of the gas phase
($\Gamma$ = 1.01).  
(c) ($\Gamma$ = 0.8) Clusters
are visible as higher intensity points in a time-averaged image, (d),
denoting slower, densely packed particles.  
(e) A portion of a collapse ($\Gamma = 0.76$).
(f) The time-averaged image shows that the particles in the collapse are 
stationary while the surrounding gas particles continue to move.
(g),(h) In a more dense system, there is an ordered phase
where all of the particles remain in motion.}
\label{fig:one}
\end{figure}


\begin{figure}
\caption{The phase diagrams for (a) N = 8000 particles and (b) N = 14,500
particles.  The filled circles denote the acceleration where the collapse
nucleates.  The open circles in (a) indicate the point where the collapse
disappears upon increasing the acceleration.  The diamonds in 
(b) show the transition to the ordered state as the acceleration is reduced.}
\label{fig:two} 
\end{figure}


\begin{figure}
\caption{Density-density correlation measurements of the granular medium.  
The measured correlations are compared to
the result from an equilibrium hard sphere Monte Carlo 
calculation [26].}
\label{fig:three} 
\end{figure}


\begin{figure} 
\caption{Probability distribution function for a single component of the
horizontal velocity on (a) linear, and (b) log scales .  
The solid line is a Gaussian distribution.  The
data is: $\circ$ - $\Gamma$ = 1.01, 
$\Box$ - $\Gamma$ = 0.80, 
$\Diamond$ - $\Gamma$ = 0.76, for N = 8k and $\nu$ = 75 Hz; 
$\triangle$ - $\Gamma$ = 1.00, N = 14.5k, $\nu$ = 90 Hz.
The large population of low speed particles is evident in (a) while (b)
shows that the tails are approximately exponential.
The data is scaled by $v_0 = (2v_{2m}^2)^{1/2}$.}
\label{fig:four} 
\end{figure}



\begin{thebibliography}{25}
\bibitem{expref}  302 stainless steel, uniformity of diameter $0.4 \%$.
N$_{max}$ = 17,275 balls constitute one monolayer at hexatic packing.
The filling fraction, $\rho$, for N particles is N/N$_{max}$.
\bibitem{stats} A Dalsa-CAD1, 128x128 pixel, 838 fps camera; (Dalsa Inc.
Waterloo, ON; CANADA).  Viewed from above, the trajectories of the particles 
are nearly straight lines between collisions.  Particle positions are analyzed
three images at a time.  To eliminate collisions, 
a momentum criterion is used, $\Delta \vec{P} < \delta$, where $\delta$ is 
a small constant.
\end{thebibliography}
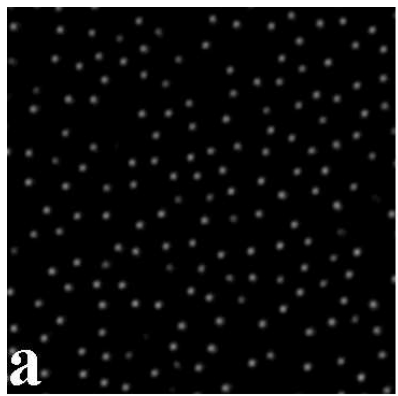
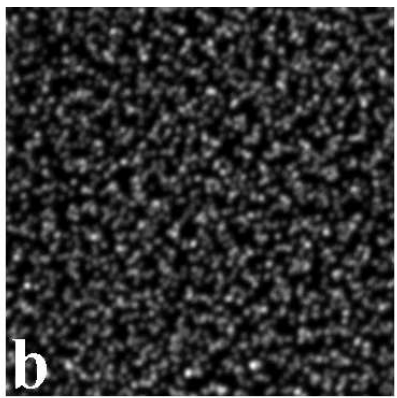
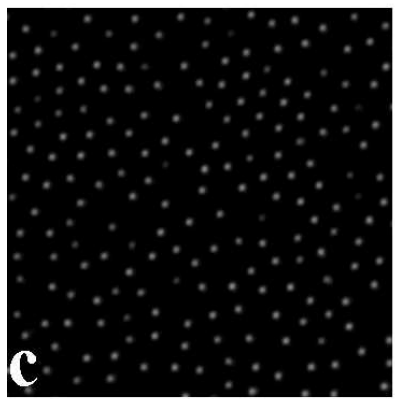
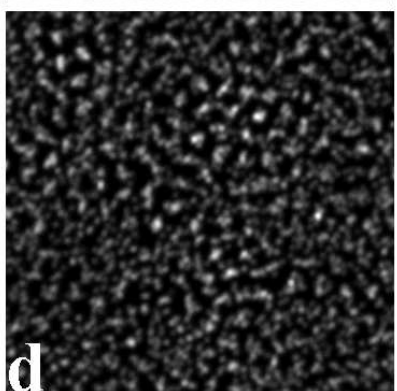
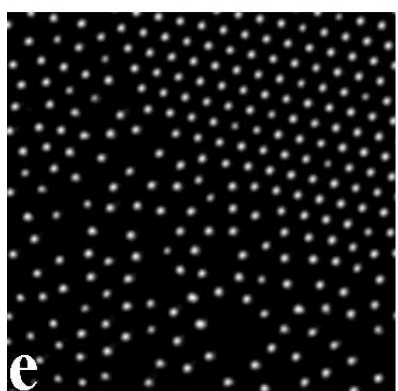
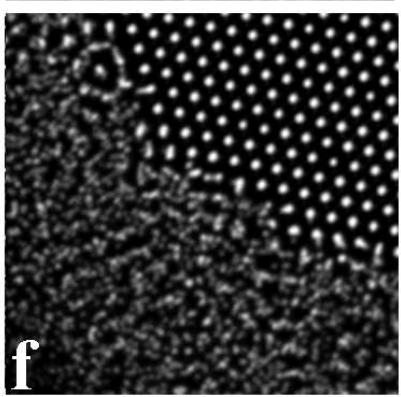
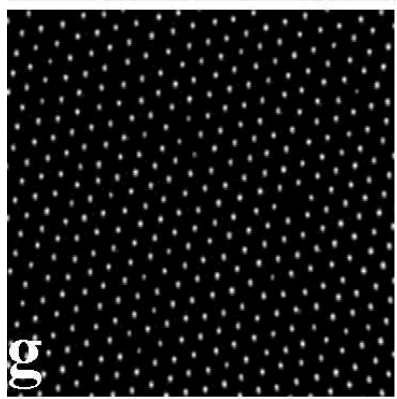
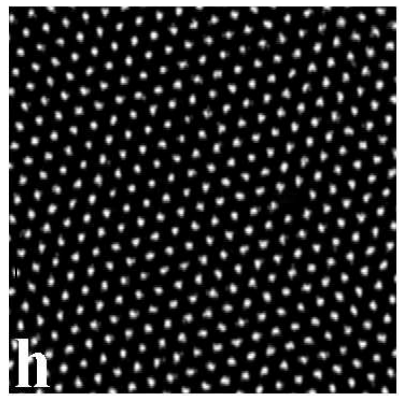
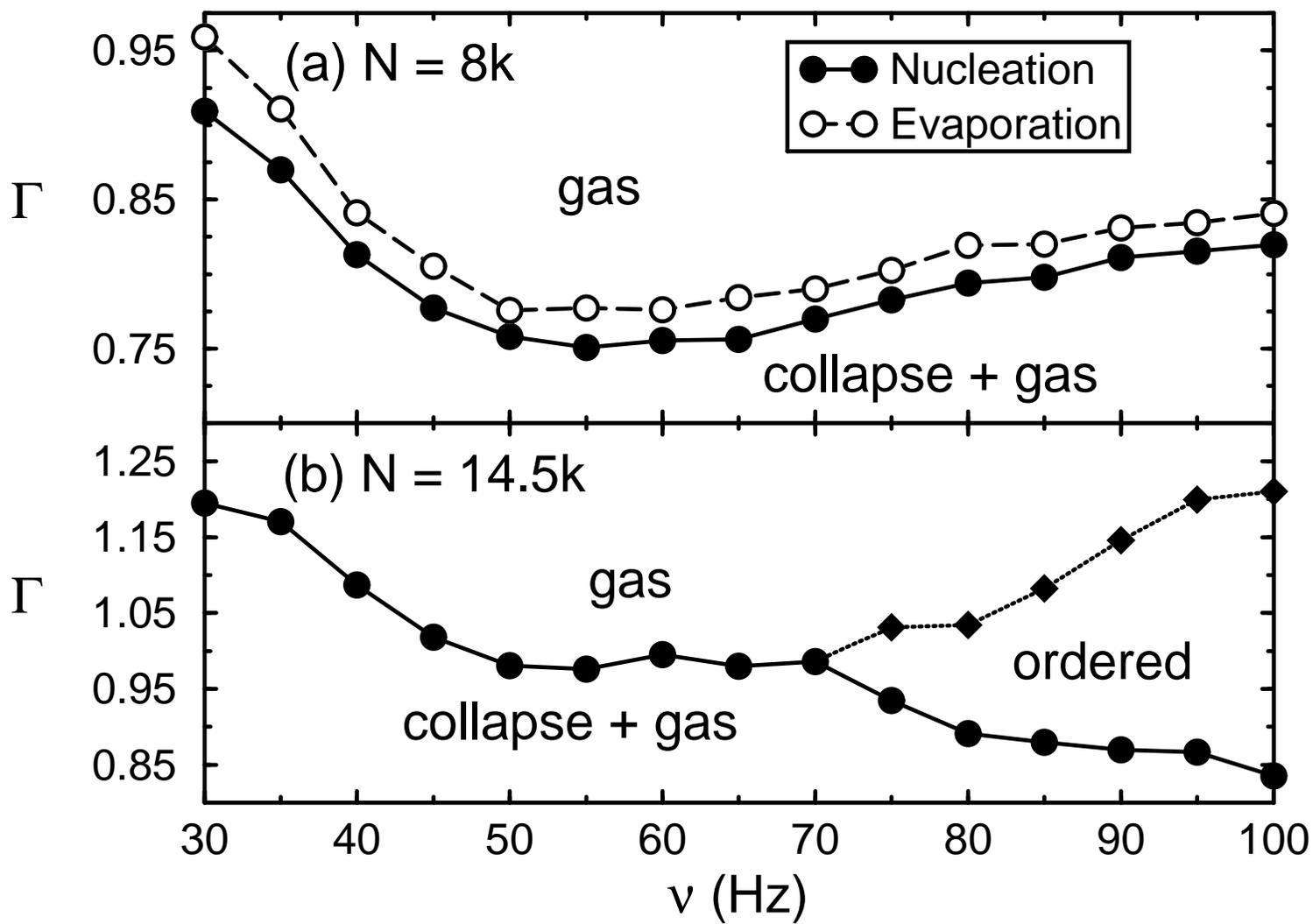
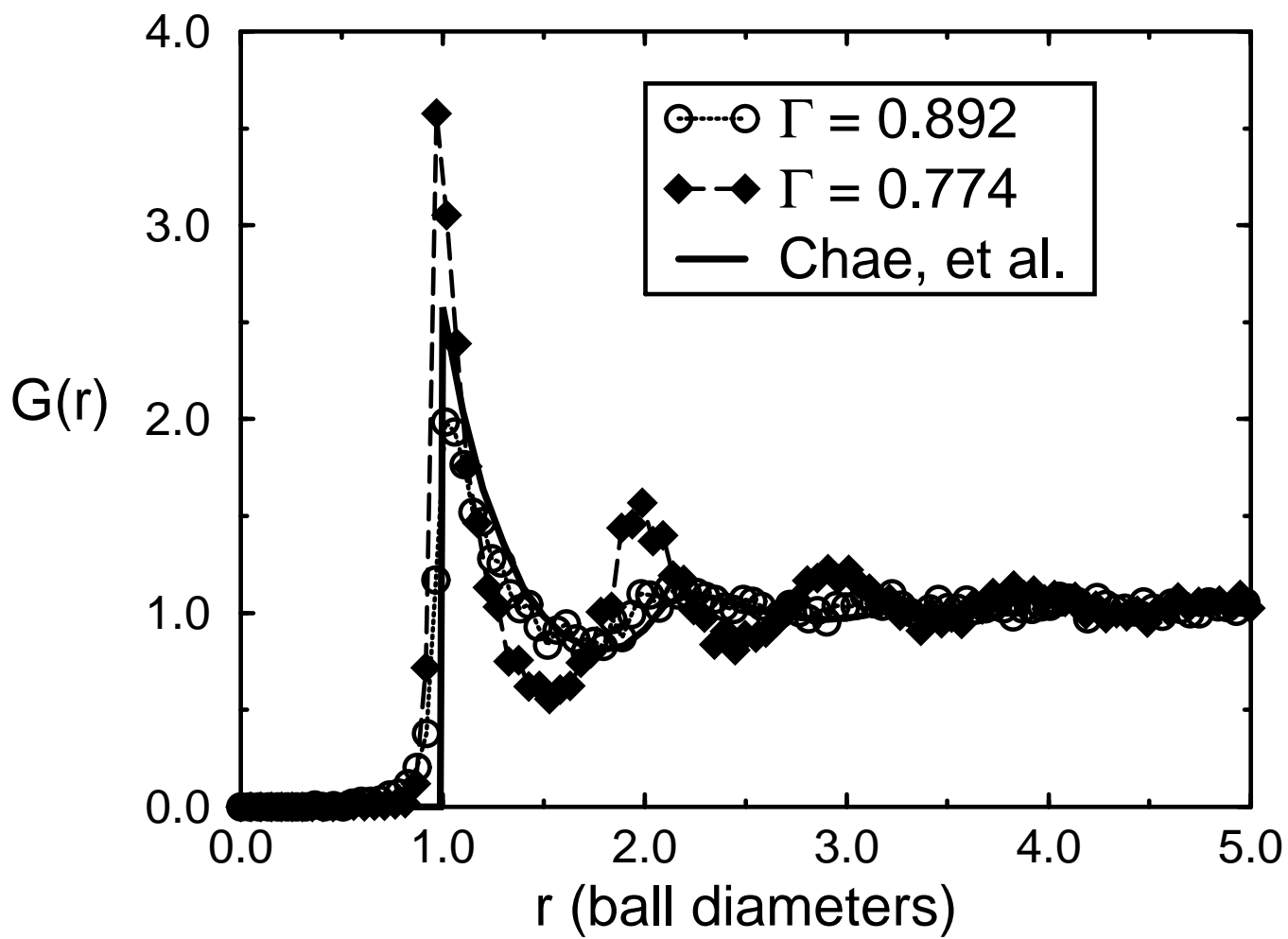
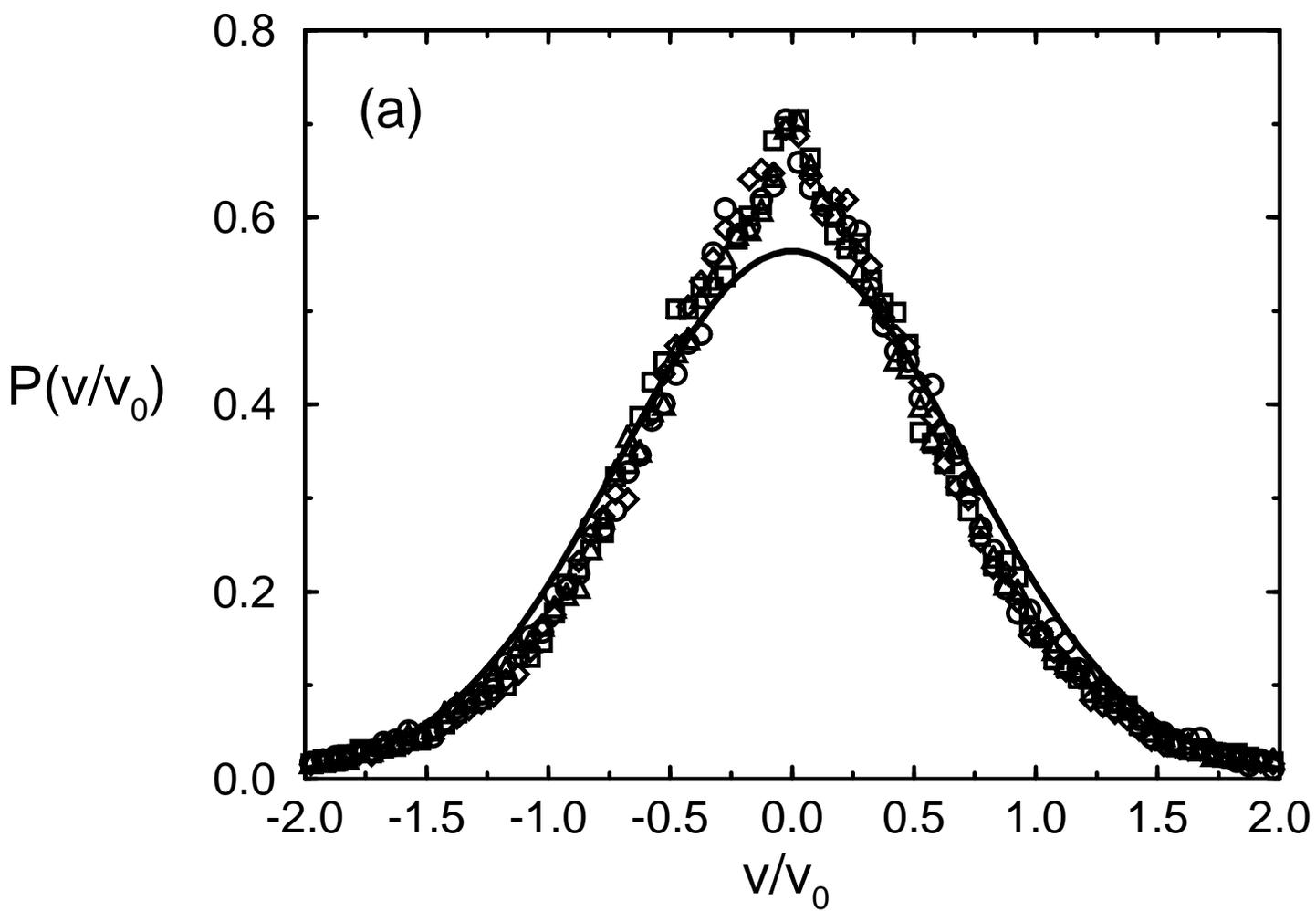
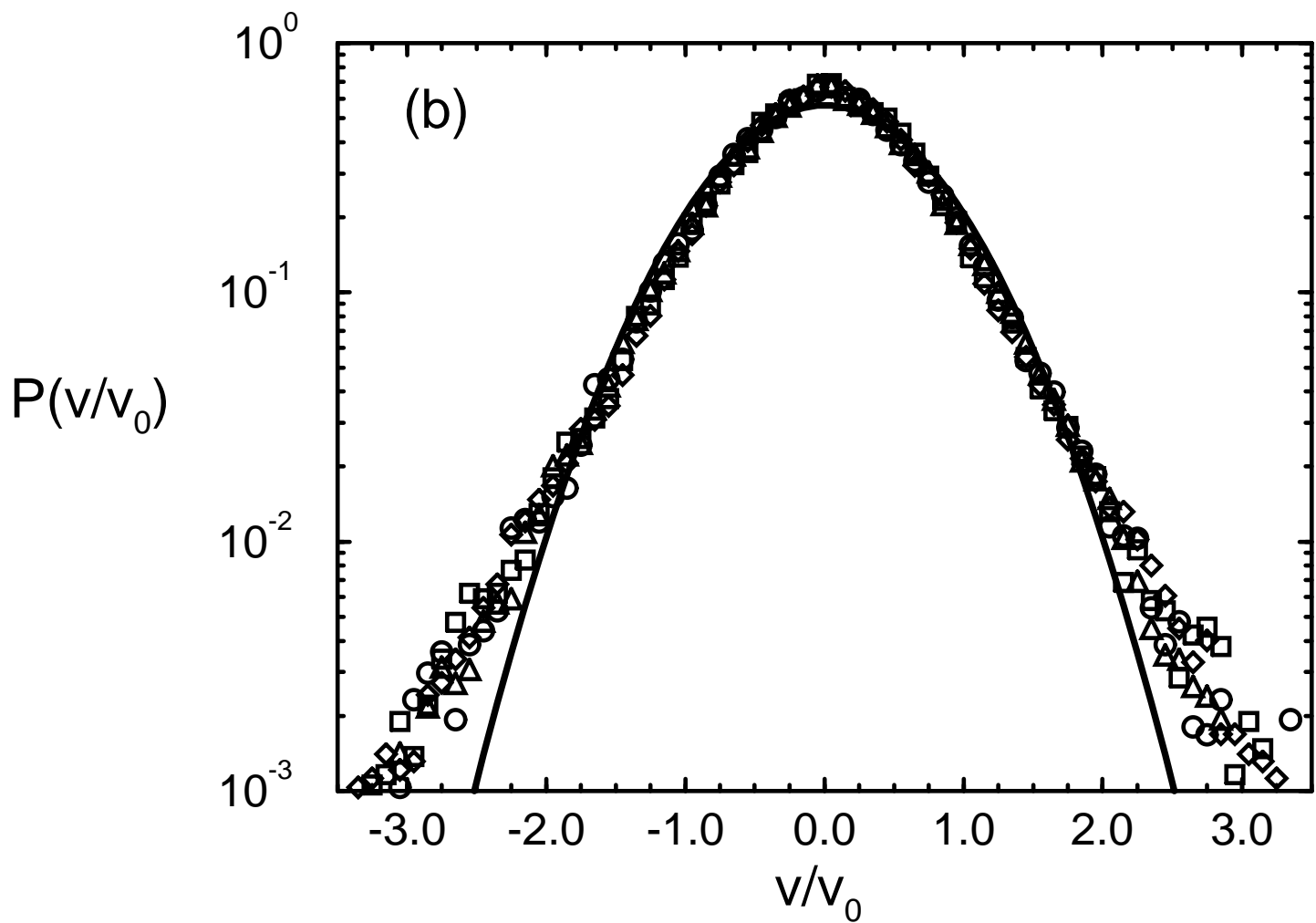
\end{document}